\documentclass[aps,prd,twocolumn,nofootinbib,superscriptaddress]{revtex4-1}
\usepackage{amsfonts}
\usepackage{amsmath}
\usepackage{amssymb}
\usepackage{bm}
\usepackage{dcolumn}
\usepackage{epsfig}
\usepackage{graphics}
\usepackage[latin1,utf8]{inputenc}
\usepackage{latexsym}
\usepackage{rotating}
\usepackage[colorlinks=true]{hyperref}
\usepackage[usenames]{color}
\usepackage{float}
\usepackage{ucs}
\usepackage{xspace} 
\usepackage{mathrsfs}
\usepackage{enumitem}
\usepackage{tabulary}
\usepackage{tabularx}
\usepackage{booktabs}
\usepackage{array}
\usepackage{adjustbox}
\usepackage[normalem]{ulem}
\usepackage{color,soul}
\usepackage{tabu}
\def\be{\begin{equation}}
\def\ee{\end{equation}}
\newcommand{\bea}{\begin{eqnarray}}
\newcommand{\eea}{\end{eqnarray}}

\newcolumntype{Y}{>{\centering\arraybackslash}X}

\newcommand{\CG}{{\mbox{\tiny CG}}}

\newcommand{\V}{{\mbox{\tiny V}}}
\newcommand{\GR}{{\mbox{\tiny GR}}}
\newcommand{\MW}{{\mbox{\tiny MW}}}
\newcommand{\s}{{\mbox{\tiny s}}}
\newcommand{\ins}{{\mbox{\tiny ins}}}
\newcommand{\intt}{{\mbox{\tiny int}}}
\newcommand{\mr}{{\mbox{\tiny mr}}}
\newcommand{\RD}{{\mbox{\tiny RD}}}
\newcommand{\peak}{{\mbox{\tiny peak}}}

\begin{document}

\title{Probing Screening and the Graviton Mass with Gravitational Waves}

\author{Scott Perkins}
\email{scottperkins2@montana.edu}
\affiliation{eXtreme Gravity Institute, Department of Physics, Montana State University, Bozeman, MT 59717, USA.}

\author{Nicol\'as Yunes}
\email{nicolas.yunes@montana.edu}
\affiliation{eXtreme Gravity Institute, Department of Physics, Montana State University, Bozeman, MT 59717, USA.}

\begin{abstract}

Gravitational waves can probe fundamental physics, leading to new constraints on the mass of the graviton. Previous tests, however, have neglected the effect of screening, which is typically present in modified theories that predict a non-zero graviton mass. We here study whether future gravitational wave observations can constrain the graviton mass when screening effects are taken into account. We first consider model-independent corrections to the propagation of gravitational waves due to screened massive graviton effects. We find that future observation can place constraints on the screening radius and the graviton mass of ${\cal{O}}(10^{2})$--${\cal{O}}(10^{4})$ Mpc and ${\cal{O}}(10^{-22})$--${\cal{O}}(10^{-26})$ eV respectively.  We also consider screening effects in two specific theories, ghost-free massive gravity and bigravity, that might not realize these types of propagation modifications, but that do provide analytic expressions for screening parameters relevant to our analysis allowing for more concrete results. However, the constraints we are able to place are small. The reason for this is that second- and third-generation detectors are sensitive to graviton masses that lead to very small screening radii in these particular models. The effect of screening, however, can become important as constraints on the graviton mass are improved through the stacking of multiple observations in the near future.  

\end{abstract}

\date{\today}

\maketitle

\section{Introduction}\label{section:intro} 

The onset of gravitational wave (GW) astrophysics has allowed new constraints of modifications to General Relativity (GR), including new bounds on the graviton mass~\cite{Abbott:2017vtc}. The latter is a particularly important quantity in a variety of fields, from quantum extensions of GR~\cite{deRham:2013qqa,Hinterbichler:2010xn,Park:2010rp,Park:2010zw} to cosmological modified theories of gravity~\cite{Chamseddine:2011bu,DAmico:2011eto,deRham:2010tw,Hinterbichler:2017sbd,Brax:2012gr,Platscher:2018voh,CLIFTON20121}. Because of this widespread interest, the graviton mass has been constrained through a host of different experiments~\cite{Finn:2001qi,Brito:2013wya,PhysRevD.9.1119,PhysRevLett.61.1159,Sakstein:2017pqi,Desai:2017dwg,Rana:2018vxn,Gupta:2018hgm}, from weak field Solar System bounds resulting in $m_g<10^{-24}$eV~\cite{Will:2018gku}, to strong field, stacked GW bounds resulting in $m_{g}<10^{-23}$eV~\cite{Abbott:2017vtc}. While Solar System constraints are currently more competitive than their GW counterparts, the former probe fifth-force type forces, while the latter constrains the propagating sector of such theories~\cite{deRham:2016nuf,Yunes:2013dva}. 

In the context of GW observations, estimated bounds on the graviton mass have used the work of Will~\cite{Will:1997bb} and extensions~\cite{Mirshekari:2011yq}. Working in the post-Newtonian (PN) framework~\cite{Blanchet:2013haa}, one first postulates a modified dispersion relation during the propagation of GWs that is based on the special-relativistic dispersion relation of massive particles. One then studies how these modifications trickle down into the observed GW phase through the stationary phase approximation~\cite{Finn:1992xs}. The end result is a correction that enters at 1PN order in the GW phase relative to the leading-order GR term and that scales with the distance traveled.  

Previous GW work on the graviton mass, however, has neglected a mechanism that is typically present in cosmological modified gravity theories: screening~\cite{VAINSHTEIN1972393,Babichev:2013usa,Babichev:2013pfa,Chkareuli:2011te}. This mechanism suppresses modifications to GR inside of some screening radius through nonlinear self-interactions of an auxiliary field, thus allowing these theories to pass Solar System tests with ease. GWs, however, typically exit this radius during their travel from the source to Earth, thus possibly acquiring modifications in their propagation \emph{only} outside the screening radius. Intuitively, one would therefore expect screening to soften constraints on the graviton mass, as its modifications to the GW phase need not always be active. Because of this softening, one should consider the effect of screening in future bounds, like those estimated in~\cite{Mirshekari:2011yq,Samajdar:2017mka,Will:2004xi,Berti:2005qd,Chamberlain:2017fjl}, to ensure modified theories are not prematurely ruled out.  

Whether screening is present in modifications to the dispersion relation of GWs depends strongly on the particular theory considered. For example, in quartic and quintic Galileon theories~\cite{Nicolis:2008in}, screening is present in the dynamics of the Galileon scalar field, but this does not percolate into the GW dispersion relation~\cite{Jimenez:2015bwa}. As such, it is not expected that this effect be present in theories like ghost-free massive gravity (dRGT)~\cite{deRham:2010kj} or bigravity~\cite{Hassan:2011zd}, which are generalizations of the Galileon family of theories~\cite{deRham:2016nuf,deRham:2014zqa}. In other theories, however, these propagation effects may be present, provided the propagation equations for the helicity-2 mode couple directly to the scalar field. 

Given this, we here take on an agnostic approach to study whether GWs can probe the graviton mass if screening is present. We first derive the correction to the GW Fourier phase when screening is present within the PN framework and using the stationary phase approximation. We find that the correction to the phase takes the same exact functional form as in the unscreened case, except that the distance parameter that enters the correction is not the usual luminosity distance, but rather a new effective quantity. This new effective distance is smaller than the luminosity distance, yielding a measure of the distance through which massive graviton modifications are active. 

We then study how well such a screened massive graviton modification could be constrained with second- and third-generation GW detectors, like aLIGO~\cite{aLIGO}, the Einstein Telescope (ET)~\cite{0264-9381-27-19-194002}, and LISA~\cite{2017arXiv170200786A}. We first modify the quasi-circular IMRPhenomD model~\cite{Husa:2015iqa,Khan:2015jqa} to construct screened massive graviton waveforms that span the whole frequency range, from inspiral to merger and ringdown. We then assume second- and third-generation detectors have detected a signal consistent with GR and predict the constraints one could place on screened modified gravity effects through a Fisher analysis~\cite{Finn:1992xs}, which should be accurate at the large signal-to-noise ratios we consider.

\begin{figure}[htb]\label{fig:example_plot}
\hspace*{-0.6cm}\includegraphics[width=9.75cm,clip=true]{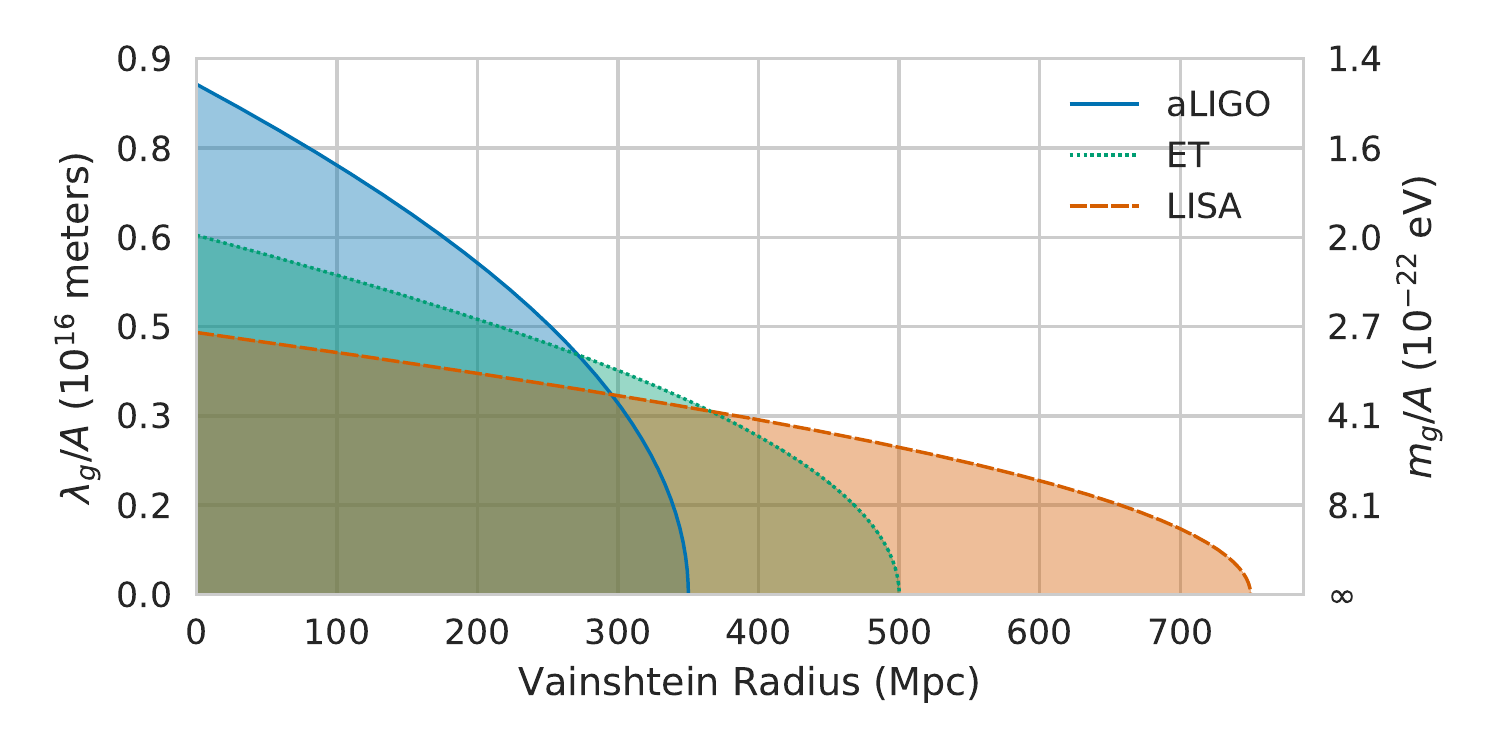}
\caption{(Color Online) Projected constraints on the graviton mass as a function of the screening radius, assuming the detection of three expected GW sources with aLIGO at design sensitivity, ET, and LISA. As the sensitivity to these screening effects varies widely for each detector, the y-axis is re-scaled appropriately through a factor $A$ presented in Table~\ref{table:1}. The effectiveness of screening becomes very apparent when the screening radius is about $D_{\tiny L}/2$, since then the constraint on $m_g$ rapidly falls to zero.}
\end{figure}

Constraints on the modifications to the GW phase lead to degenerate constraints on the graviton mass and the screening radius, when these parameters are treated as independent. Figure~\ref{fig:example_plot} shows these degenerate constraints, where the shaded regions would be disallowed, given GW observations consistent with GR and produced by the systems in Table~\ref{table:1}. When the screening radius is small, then constraints on the graviton mass reduce to previous estimates~\cite{Mirshekari:2011yq,Samajdar:2017mka,Will:2004xi,Berti:2005qd,Chamberlain:2017fjl}. When the screening radius is large, the projected constraints deteriorate, until for sufficiently large radii no constraints on the graviton mass are possible any longer. The latter occurs when the screening radius is roughly half the luminosity distance to the source, as then the graviton mass effects are completely screened. 

\begin{table}[htb]
\begin{tabularx}{8.5cm}{>{\centering\arraybackslash}X | >{\centering\arraybackslash}X >{\centering\arraybackslash}X >{\centering\arraybackslash}X >{\centering\arraybackslash}X}
Detector & $m_1$ ($M_\odot$) & $m_2$ ($M_\odot$)& $D_{\tiny L}$ (Mpc) & $A$ \\ \hline \hline
aLIGO & 20&15 &700 &1\\ \hline
ET & 35 &10& $10^3$&10\\ \hline
LISA & $10^6$& $10^5$ & $1.5\times 10^3$ & $10^4$ \\ 
\end{tabularx}
\caption{Properties of the binary systems considered in Fig.~\ref{fig:example_plot} to estimate projected bounds on the graviton mass and the screening radius. The parameter $A$ is a factor used to re-scale the y-axis of the figure.}
\label{table:1}
\end{table}

We conclude with a study of projected constraints on screening and the graviton mass in specific cosmological modified gravity theories.  In particular, we consider two massive gravity theories, dRGT and bigravity, in which the screening radius is a function of the graviton mass~\cite{deRham:2014zqa,Babichev:2013usa}. The effect of screening on the scalar field in these theories does not appear to affect the propagation of the tensor modes~\cite{deRham:2016nuf,deRham:2014zqa}, but these models give explicit functional forms for the screening radius as a function of the graviton mass. This allows for a more tangible result that might be indicative of what one may expect in other more general theories that do include screening of the tensor modes. In these theories, however, the scaling of the screening radius is such that the graviton mass that second- and third-generation detectors are sensitive to lead to tiny screening effects. Only once constraints on the graviton mass become of ${\cal{O}}(H_{0})$ do screening effects begin to become important in these theories. Such graviton mass constraints, however, could be achievable through the stacking of multiple observations with third-generation detectors~\cite{Ghosh:2016qgn}.     

The remainder of this paper deals with the details of the calculations that led to the results summarized above. 
Section~\ref{sec:GW-basics} presents the basics of screening and its impact on the GW response function. 
Section~\ref{sec:Fisher} describes the Fisher analysis methodology that we employ to estimate constraints. 
Section~\ref{sec:proj-const} presents the projected constraints we obtained. 
Section~\ref{sec:conclusions} concludes and points to future research. 
Henceforth, we follow mostly the conventions of~\cite{Misner:1974qy}, where Greek letters represent spacetime indices, Latin letters represent parameter indices, and for the most part we employ geometric units in which $G = 1 = c$. 

\section{Gravitational Wave Propagation in Screened Theories}
\label{sec:GW-basics}

In this section, we calculate how the propagation of GWs is affected by a screening radius in theories with a massive graviton. 
We begin by describing the basics of Vainshtein screening, therefore setting notation. 
We then continue with the calculation of the modifications to the propagation speed of GWs, and how this percolates 
to the Fourier transform of the response function in the stationary phase approximation.  

\subsection{Vainshtein Screening}
Vainshtein screening is the process by which nonlinear interaction in the field equations suppress modifications to GR in a certain regime of spacetime. Let us here outline the basic foundation behind Vainshtein screening; a more in depth description can be found in several reviews on the topic, e.g.~\cite{Babichev:2013usa,deRham:2014zqa,Hinterbichler:2011tt,Schmidt-May:2015vnx} or in works like~\cite{Belikov:2012xp,Babichev:2012re,Koyama:2011yg,Gannouji:2011qz,Kase:2013uja}. 

First discovered in 1972 by Vainshtein~\cite{VAINSHTEIN1972393}, the idea of screening has seen waves of interest, with a renewed popularity recently due to recent work on theories like dRGT and bigravity. Originally, the mechanism was introduced as the solution to the van Dam-Veltman-Zakharov discontinuity~\cite{Zakharov:1970cc}, a problem that had plagued massive gravity in the 1970's. This discontinuity arises as an inability of massive gravity to match (at linear order) GR solutions in the limit of a vanishing graviton mass, resulting in easily measured discrepancies between predictions and observations in the Solar System and gravitational lensing experiments. However, Vainshtein noticed that these deviations came at  linear order in the metric perturbation, and that by including higher order terms, these modifications could be suppressed around massive sources, reviving the viability of massive gravity theories~\cite{Porrati:2002cp,VAINSHTEIN1972393}. 

Later on, a different issue, the so-called Boulware-Deser ghost~\cite{PhysRevD.6.3368}, began to haunt massive gravity theories. This unbounded-from-below sixth degree of freedom again muted interest for a number of years until de Rham, Gabadaze, and Tolley formulated dRGT to specifically eradicate this unphysical degree of freedom~\cite{deRham:2010kj}. This gave rise to renewed interest in massive gravity theories and spawned bigravity as an extension, which also exhibits Vainshtein screening effects~\cite{Babichev:2013pfa}. Since then, the Vainshtein mechanism has been crucial in keeping theories viable in the wake of highly accurate Solar System tests, as it has been shown to be effective at suppressing ``fifth force'' modifications to the Newtonian potential~\cite{deRham:2014zqa}.

Screening, however, is usually analyzed in the static, spherically symmetric case. It remains a bit of an open topic of research whether the screening persists within dynamical systems for massive gravity, and how effective the screening actually is. Recently, the mechanism has been examined in the context of the cubic Galileon theories (a simpler, but related theory to dRGT) numerically and it has been shown that the energy loss of binary systems through scalar radiation is indeed suppressed (albeit not to the full extent of the suppression of the fifth force in static scenarios)~\cite{Dar:2018dra}. 

To illustrate the effect of screening in a simple scenario, let us focus on a static, spherically-symmetric scenario in the context of the cubic Galileon theories. Following~\cite{deRham:2014zqa}, the cubic Galileon action in geometric units is
\begin{align}\label{eq:cgaction}
\nonumber S_{\CG} &= \int d^4 x \sqrt{-g} \left[ \frac{R}{16 \pi } - \frac{1}{2} \partial_\mu \phi \partial^\mu \phi \right. \\
&\left. -\frac{1}{\Lambda^3} \partial_\mu \phi \partial^\mu \phi \Box \phi +\sqrt{8 \pi} \phi T  \right] + S_m[g]\,,
\end{align}
where $\Lambda=[m_g^2 /(\hbar^2 \sqrt{8\pi})]^{1/3}$, the Ricci scalar $R$ is associated with the metric tensor $g_{\mu \nu}$, the scalar field $\phi$ couples to the trace of the matter stress-energy tensor $T$ and the metric tensor couples to matter through $S_{m}$.

Varying Eq.~\eqref{eq:cgaction} in a static and spherically symmetric system with a stress energy tensor trace $T = -M \delta(r)/4\pi r^2$ results in the equation of motion of the scalar field
\begin{equation} \label{eq:EOM-vains1}
\frac{1}{r^2}\frac{d}{dr}\left[r^2 \frac{d\phi}{dr} + \frac{r}{\Lambda^3}\left(\frac{d\phi}{dr}\right)^2\right] = \frac{ M}{\sqrt{2 \pi} }\frac{\delta(r)}{r^2}\,,
\end{equation}
which, after direct integration of the first derivative, gives 
\begin{equation}\label{eq:cg_vainshtein}
\frac{1}{r}\frac{d\phi}{dr} + \frac{1}{r^2\Lambda^3}\left(\frac{d\phi}{dr}\right)^2 = \frac{ M}{\sqrt{2 \pi}} \frac{1}{r^3}\,.
\end{equation}

The importance of Vainshtein screening is illustrated in Eq.~\eqref{eq:cg_vainshtein}. With the definition 
\begin{equation}\label{eq:rv_cg}
r_\V = \frac{1}{\Lambda} \left( \frac{M}{\sqrt{2 \pi } }\right)^{1/3}\,,
\end{equation}
the separation of scales is apparent, and allows us to approximate the gradient of the scalar field from Eq.~\eqref{eq:cg_vainshtein} via
\begin{equation}
\frac{d \phi}{dr} \approx \begin{cases}
							\frac{M}{\sqrt{2 \pi}}\frac{1}{r^2} & r \gg r_\V\,,\\
							\frac{M}{\sqrt{2 \pi}}\frac{1}{r^2} (\frac{r}{r_\V})^{3/2} & r \ll r_\V\,.
						\end{cases}
\end{equation}
For regimes of spacetime $r \ll r_{\V}$, the evolution of the scalar field is suppressed by the $(r/r_\V)^{3/2}$ factor, which then suppresses the ``fifth force'' modifications this scalar field would introduce. 

For dynamical systems, the situation is slightly more complicated. Let us leave cubic Galileon theories and consider instead the generic action 
\begin{align}\label{eq:generic_action}\hspace*{-.5cm}
S &= \int d^4 x \frac{\sqrt{-g}}{8 \pi }\left[ \frac{R}{2} -\frac{1}{2}\partial_\mu \phi \partial^\mu \phi 
\right. 
\nonumber \\
&\left.
-\mathcal{U}(g,\phi,\partial \phi) + \mathcal{L}_{\text{matter}}(g,\phi) \right]\,,
\end{align} 
where $\mathcal{U}$ is a kinetic self-interaction term that depends on the metric, as well as derivatives of the auxiliary field. The field equations associated with this action are schematically
\begin{subequations}
\begin{equation}\label{eq:generic_metriceom}
G_{\mu \nu} \propto T_{\mu \nu} -\partial_\mu\phi \partial_\nu \phi - \frac{\delta \mathcal{U}}{\delta g_{\mu\nu}}\,,
\end{equation}
\begin{equation}\label{eq:generic_scalareom}
\Box \phi + \frac{\delta \mathcal{U}}{\delta \phi} \propto \frac{\delta\mathcal{L}_{\text{matter}}}{\delta \phi}\,.
\end{equation}
\end{subequations}
The equation of motion for the scalar field resembles that in Eq.~\eqref{eq:EOM-vains1}, with the identification of the variation of $\mathcal{U}$ in Eq.~\eqref{eq:generic_scalareom} with the non-linear term in $\phi$ in Eq.~\eqref{eq:EOM-vains1}. 

The analysis of the evolution of the scalar field now follows closely the analysis of Eq.~\eqref{eq:EOM-vains1}. If the kinetic, self-interaction term of Eq.~\eqref{eq:generic_scalareom} dominates through the Vainshtein mechanism, (a) the scalar mode is not excited because of a suppression of its coupling to matter (verified numerically in~\cite{Dar:2018dra}), and (b) the impact of the scalar field on the metric perturbation in the radiation regime, but still in the Vainshtein region, is also suppressed. This implies that GR is recovered within the Vainshtein radius, even in terms of propagation effects. Outside the Vainshtein radius, the self-interaction term in Eq.~\eqref{eq:generic_scalareom} becomes much smaller than $\delta\mathcal{L}_{\text{matter}}/ \delta \phi-\Box \phi$, and $\phi$ is allowed to grow, in turn allowing $\delta \mathcal{U}/\delta g_{\mu\nu}$ to affect the propagation of the GW tensor modes. 

Before proceeding, let us define the screening radius for massive objects that we will use later. This quantity differs from theory to theory, but as an example, the screening radius for a galaxy with Schwarzschild radius $r_s$ in dRGT and bigravity takes the form~\cite{deRham:2014zqa} s
\begin{equation}
\label{eq:drgtradius}
r_\V = r_s^{1/3} \lambdabar_g^{2/3}\,,
\end{equation}
whereas the screening radius in non-linear Fierz-Pauli gravity with a general potential is~\cite{Babichev:2013usa}
\begin{equation}
\label{eq:general_massive_radius}
r_\V =  r_s^{1/5}  \lambdabar_g^{4/5}\,,
\end{equation}
with the reduced Compton wavelength $\lambdabar_g = \hbar/m_g$.

To remain agnostic throughout most of the rest of the paper, the Vainshtein radius will be kept as an independent variable $r_\V$, only studying specific models close to the end. 

\subsection{Modifications to GW Propagation Speed}\label{wavespeed}

Let us focus on the modifications to the GW observable introduced by a massive dispersion relation with screening effects. The modifications without screening were originally derived by Will~\cite{Will:1997bb}, and then extended in~\cite{Mirshekari:2011yq} to consider more generic dispersion relations. We begin by reviewing Will's original derivation and then incorporating screening effects. 

Let us then postulate that a graviton with a non-zero mass obeys the generic dispersion relation of a massive particle 
\begin{equation}
\label{eq:SRdisp}
E^2 = p^2 + m_g^2 \,.
\end{equation}
Recasting the energy and momentum of the graviton as $E = \hbar \omega$ and $p = \hbar k$, the \emph{classical} or \emph{particle} speed of propagation is the same as the group velocity for the graviton, namely 
\begin{equation}
\label{eq:gravspeed}
\left(\frac{d \omega}{dk} \right)^2= v_g^2 = 1 - \frac{m_g^2}{E^2} \,.
\end{equation}

Equation \eqref{eq:gravspeed} is where we incorporate additional modifications induced by screening. More specifically, let us consider a modified theory of gravity that excites extra degrees of freedom beyond the two of GR. When these extra degrees of freedom are suppressed, the modified theory reduces essentially to GR, while when they are excited modifications are present. Moreover, let us assume that the extra degrees of freedom are non-linearly coupled, such that they are suppressed inside some screening radius, but not suppressed at larger radii in intergalactic space. The speed of propagation of gravitons is then modified from Eq.~\eqref{eq:gravspeed} into 
\begin{subequations}
\label{eq:disp_screen}
\begin{equation}
v_g^2 = 1 - \frac{m_g^2 }{E^2} \Theta(r-r_{\V,h}) \Theta(D_{\V,\MW} -r)\,,
\end{equation}
\begin{equation}
D_{\V,\MW}=D_L -r_{\V,\MW}\,,
\end{equation}
\end{subequations}
where $r_{\V,h}$ and $r_{\V,\MW}$ are the screening radii of the host galaxy and the Milky Way respectively, $D_L$ is the luminosity distance, and the coordinate $r$ is measured from the center of mass of the source. Although Eq.~\eqref{eq:disp_screen} is purely phenomenological, it is still interesting to consider what such screening effects would do to constraints on the graviton mass. 

A bound for the Compton wavelength of the graviton purely from the difference in travel time between a photon and graviton can quickly be derived. Noting that the Compton wavelength $\lambda_g = h/m_g $, we have that
\begin{equation}
\begin{aligned}
\label{eq:DTOA}
\Delta t = \Delta t_a - & (1+z) \Delta t_e \approx D \left( 1-v_g\right)\,, \\
\Rightarrow \lambda_g &= \sqrt{\frac{D}{2 f^2 \Delta t}}\,,
\end{aligned}
\end{equation}
with $D$ some distance measure of \emph{unscreened} space between the source and the detector and $f$ being the graviton's frequency. 

\begin{figure*}
\includegraphics[width=.9\textwidth]{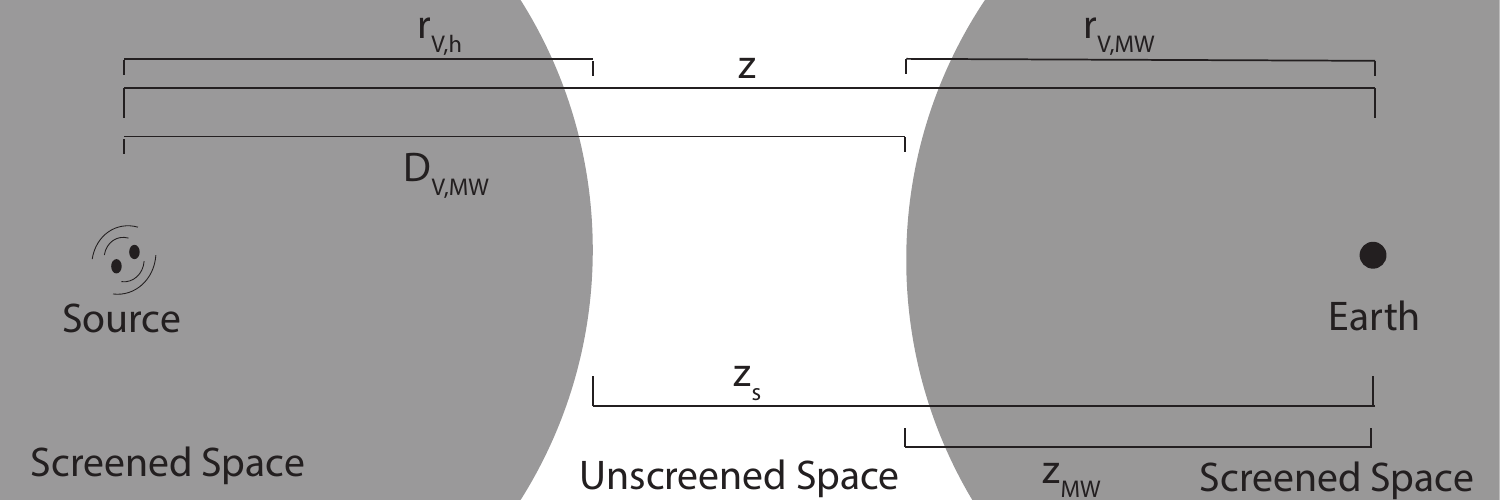}
\caption{Schematic of the binary/detector system with various distances labeled for clarity. $z$ is the redshift to the source, and $z_\s$ and $z_{\MW}$ are the redshifts to the edges of the screening effects for the host galaxy and the Milky Way, respectively. $r_{\V,h}$ ($r_{\V,\MW}$) is the Vainshtein radius for the host galaxy (Milky Way). $D_{\V,\MW}$ is the distance from the source to the edge of the Milky Way Vainshtein regime, equal to the luminosity distance to the source minus the Vainshtein radius of the Milky Way ($D_L - r_{\V,\MW}$)}
\label{fig:sys_schem}
\end{figure*}

A more rigorous calculation is required to figure out exactly what this unscreened distance is. Let us then compute the coordinate distance traveled by a massive graviton in a FLRW universe with metric 
\begin{equation}
\label{eq:metric}
ds^2 = -dt^2 + a(t)^2\left[d\chi^2 + \Sigma(\chi)^2 d\Omega^2\right]\,.
\end{equation} 
The radial momentum $p_\chi = a(t_e)(E_e^2 - m_g^2)^{1/2}$ is conserved because it is normalized via $p_{\mu} p^{\mu} = - m^{2}_{g}$, where $t_{e}$ is the emission time. Since the classical or particle velocity of the graviton $d\chi/dt = p^{\chi}/p^{t}$, we then have that 
\begin{align}
\label{eq:gravdist}
\frac{d\chi}{dt} = \frac{1}{a} \left(1+ \frac{a^2 m_g(r)^2 }{p_\chi^2}\right)^{-1/2}\,, 
\end{align}
where $m_g(r)^2 = m_g^2 \Theta(r-r_{\V,h}) \Theta(D_{\V,\MW} -r)$. This differential equation can be expanded in the small parameter $m_g(r)/E_e \ll 1$, and then solved to find the coordinate distance at emission $\chi_{e}$, namely
\begin{align}
\chi_e &\approx \int_{t_e}^{t_a}\frac{dt}{a(t)} - \frac{1}{2}\frac{m_g^2}{a^2(t_e) E_e^2}
\nonumber \\
& \times \int_{t_e}^{t_a} a(t)\Theta(r-r_{\V,h}) \Theta(D_{\V,\MW} -r) dt 
\nonumber \\
&= t_a - (1+z)t_e -\frac{1+z}{2}\frac{m_g^2}{ E_e^2} {\cal{D}}\,,
\end{align} 
where we have used that $\Delta t_e << H^{-1}$, $1+z = a_0/a_e$ and ${\cal{D}}$ is a new cosmological distance measure defined by 
\begin{align}
\label{eq:defD}
\frac{{\cal{D}}}{1+z} &= \int_{t_e}^{t_a} a(t) \Theta(r-r_{\V,h}) \Theta(D_{\V,\MW} -r) dt
\nonumber \\
&= \int_{0}^{z_\s} \!\!\! \frac{1}{H(z')}  \frac{dz'}{(1+z')^2} - \int_{0}^{z_{\MW}} \!\!\!\!\!\!\frac{1}{H(z')}  \frac{dz'}{(1+z')^2} \,,
\end{align}
where $z_\s$ and $z_{\MW}$ represent the redshift of the edge of the screening effects for the source galaxy and the Milky Way respectively, while $H$ is the Hubble parameter, which for a $\Lambda$CDM universe at late times is $H \approx H_0\sqrt{\Omega_m(1+z)^3 + \Omega_\Lambda}$, with $H_{0} = 67.31$ km/s/Mpc the Hubble constant, $\Omega_{m} = 0.315$ the matter density and $\Omega_{\Lambda} = 0.685$ the dark energy density~\cite{Ade:2015xua}. Figure \ref{fig:sys_schem} shows a schematic view of the system with the quantities labeled for clarity. When screening is absent, $z_{\s} \to z$ and $z_{\MW} \to 0$, rending Eq.~\eqref{eq:defD} identical to Eq.~$(2.5)$ in~\cite{Will:1997bb}. The effect of the screening corrections is to ``soften'' the effect of a massive graviton, hiding the mass for stretches of the propagation distance.

We can now compare the coordinate distances traveled by a massive graviton and a massless photon between Earth and the source. The coordinates describe the same location, so the difference gives
\begin{equation}
\label{eq:DTOAcosmo}
\chi_{g} - \chi_{\rm null} = 0 = \Delta t -  \frac{(1+z)}{2}\frac{m_g^2}{E_e^2}{\cal{D}}\,. 
\end{equation}
Given a observed time delay $\Delta t$ between an impinging gravitational wave and an impinging electromagnetic wave, one could then place the constraint
\begin{equation}
\label{eq:lamBoundTOA}
\lambda_g \geq \sqrt{\frac{(1+z){\cal{D}}}{2\Delta t f^2}}\,,
\end{equation}
where ${\cal{D}}$ is defined in Eq.~\eqref{eq:defD}. This expression for the constraint on $\lambda_{g}$ is equivalent to the one obtained using Eq.~\eqref{eq:DTOA}, where we now see that the \emph{unscreened} distance $D$ appearing in that equation is really $(1+z) {\cal{D}}$. 

\subsection{Phase Modifications from Finite $\lambda_g$}
While the above calculations illustrate the effectiveness of constraining the mass of the graviton through direct comparison of GW and photon travel times, more stringent constraints can be derived by examining the phase modifications of a gravitational signal coming from coalescing binaries. This phase modification has already been investigated using the GW150914 signal, which has led to the constraint $\lambda_g \gtrsim 10^{13} $ km at 90\% confidence through matched filtering~\cite{PhysRevLett.116.221101,Ellis:2016rrr} and later a bound of $\lambda_g>1.6\times 10^{13}$km by combining several observations~\cite{Abbott:2017vtc} (assuming no screening effects are present). 
For different constraints on the graviton mass using the GW150914 observation, see{~\cite{Bicudo:2016pps}}. In that work, the authors were concerned with constraints on the screening of the \emph{Newtonian potential},  as opposed to the screening of modifications to GR analyzed in this work.

Let us then derive the modifications to the phase when screening is present. Emulating the derivation of Will~\cite{Will:1997bb}, the coordinate difference calculation of Sec.~\ref{wavespeed} can be applied to two gravitons with different energies $E$ and $E'$, instead of a photon and a graviton. The difference in travel time between two gravitons is then useful for deriving the phase modifications to a gravitational wave signal. The resulting difference in travel time is then given by
\begin{equation}
\label{eq:DTOAgrav}
 \Delta t_a = (1+z)\left[ \Delta t_e + \frac{\cal{D}}{2 \lambda_g^2}\left(\frac{1}{f_e^2}-\frac{1}{{f'}_e^2}\right)\right]\,.
\end{equation}
We see that this equation is identical to that derived in~\cite{Will:1997bb} except that here the quantity ${\cal{D}}$ contains screening modifications. 

The later observation then allows us to immediately write down the modification to the GW Fourier phase. Using the restricted PN approximation, the Fourier transform of the response function for a GW emitted by a compact binary is simply 
\begin{subequations}\label{eq:pnwaveform}
\begin{equation}
\tilde{h}(f) = A({f}) e^{i \Psi(f)} \,,
\end{equation}
where the Fourier amplitude is
\begin{equation}
A({f}) = \sqrt{\frac{\pi}{30}} \frac{\mathcal{M}_z^2}{D_L} u^{-7/6} \,,
\end{equation}
and the Fourier phase is
\begin{align}
\label{eq:PNphase}
\Psi({f}) &= 2 \pi {f} {t}_c - \phi_c - \pi/4 + \frac{3}{128} u^{-5/3} - \beta u^{-1} \nonumber \\
& +\frac{5}{96}\left( \frac{743}{336} + \frac{11}{4} \eta\right) \eta^{-2/5}u^{-1} - \frac{3 \pi }{8} \eta^{-3/5}u^{-2/3} \,,
\end{align}
\end{subequations}
with $u=\pi \mathcal{M}_z\tilde{f}$ and $\mathcal{M}_z= (1+z)\eta^{3/5} m$ is the redshifted chirp mass. The Fourier phase is presented here to $1.5$PN order, but it is straightforward to include the higher PN order terms that have already been calculated (see e.g.~the references in \cite{Husa:2015iqa,Khan:2015jqa}).  

The Fourier phase clearly depends on quantities related to the binary itself, like the redshifted chirp mass ${\cal{M}}_z$, with $\eta = m_{1} m_{2}/m^{2}$ the symmetric mass ratio and $m=m_{1}+m_{2}$ the total mass, as well as the time and phase of coalescence $t_{c}$ and $\phi_{c}$. But this phase also depends on massive graviton modifications, encoded in the parameter $\beta$, which is given by
\begin{equation}\label{eq:beta_relation}
\beta = \frac{\pi^2 {\cal{D}} \mathcal{M}_z}{\lambda_g^2 (1+z)} \,,
\end{equation}
where $\cal{D}$ is defined as in Eq.~\eqref{eq:defD}. Clearly then, the modifications to the Fourier phase are identical to those of a massive graviton, except that the quantity ${\cal{D}}$ is an effective luminosity distance that accounts for screening effects.

\section{Fisher Analysis and Computational Framework}
\label{sec:Fisher}

In the next section, we outline the details behind our model creation and statistical analysis. We briefly introduce the basic foundations of a Fisher analysis and define the calculations used to find the estimated bounds on our model parameters. We then go on to describe the algorithm chosen to model our sample waveforms, and finally, we present the parameters of the models we used in this study.
\subsection{Fisher Information Matrix} \label{subsect:fisher}
The accuracy to which parameters $\lambda^{a}$ in a model (the waveform template in this case) can be estimated from a data set (the GW signal in this case) can be approximated through the Fisher information matrix~\cite{Yunes:2013dva} (referred to as the Fisher matrix) in the high signal-to-noise ratio (SNR) limit~\cite{PhysRevD.77.042001,PhysRevLett.107.191104}. The Fisher matrix is defined via 
\begin{equation}
\label{eq:fisherdef}
\Gamma_{ab} = (\partial_a h| \partial_b h) \,,
\end{equation}
where the inner product is
\begin{equation}
\label{eq:innerproddef}
(h_1 | h_2) = 2\int \frac{h_1 h_2^*+h_2h_1^*}{S(f)}df \,,
\end{equation}
with $S(f)$ the sensitivity noise density of the detector and the star representing complex conjugation. The inner product also leads to the definition of the SNR via
\begin{equation}
\rho^2 = ( h| h) \,.
\end{equation}

For computing the Fisher elements, we use the projected noise density curves of two ground-based detectors, advanced LIGO at design sensitivity~\cite{aLIGO} and the ET (or specifically ET-D)~\cite{0264-9381-27-19-194002}, a proposed third-generation detector. ET is projected to improve drastically upon aLIGO (by a factor of 10 to 100, especially at low frequencies) and is expected to come online in the 2030s. For space-based detectors, we will use the final LISA design submitted by the ESA~\cite{2017arXiv170200786A}. This design will consist of six-links with 2.5 Gm arms and low acceleration noise demonstrated possible with LISA Pathfinder~\cite{PhysRevLett.116.231101}. 

The calculation of the Fisher matrix requires the evaluation of derivatives of the waveform model and then the integration of these derivatives normalized by the spectral noise density [see e.g.~Eq.~\eqref{eq:innerproddef}]. The integration for each element of the Fisher matrix is here truncated at the frequency for which the signal is about a tenth of the noise spectrum. If the model was based on a system that contained a neutron star, the cut off frequency was instead set to an approximation of the frequency at contact $f_{\rm contact} =(M c^3/\pi^2 (24 \text{ km})^3)^{1/2}$. This cutoff was chosen to ensure the accuracy of the model, as this frequency is a conservative estimate of the point at which the faithfulness of our waveforms breakdown.

Given the Fisher matrix, the variance of any estimated model parameter $\hat\lambda^{a}$ can then be approximated from the Cramer-Rao bound
\begin{equation}\label{eq:inversefisher}
\sigma_{\lambda^{a}} \geq \sqrt{\Sigma^{aa}} \,,
\end{equation}
where $\Sigma^{ab} = (\Gamma_{ab})^{-1}$ is the variance-covariance matrix, and no sum over the index $a$ is here implied. In this paper, we will assume a GW signal consistent with GR at a sufficiently high SNR has been detected; we will synthesize this injection through a waveform model evaluated at $\beta = 0$. We will then estimate the accuracy to which the parameters of a waveform model, including $\beta$, can be estimated using a Fisher matrix approach.   

Given an estimated bound on $\beta$, we can then calculate a projected bound on the physical constants that $\beta$ depends on. Simple propagation of error gives
\begin{align}\label{eq:error_prop}
\sigma_{\bar{m}_{g}^{2}}^{2} &= 
\left(\frac{\partial \bar{m}_{g}^{2}}{\partial \beta}\right)^{2} \sigma_{\beta}^{2} + 
\left(\frac{\partial \bar{m}_{g}^{2}}{\partial {\cal{M}}}\right)^{2} \sigma_{\cal{M}}^{2}  + \left(\frac{\partial \bar{m}_{g}^{2}}{\partial D_L}\right)^{2} \sigma_{D_L}^{2}
\nonumber \\
&+ \sum_{i \neq j \in \{\beta,\mathcal{M}_z,D_L\}} 2\frac{\partial \bar{m}_g^2}{\partial \lambda_i} \frac{\partial \bar{m}_g^2}{\partial \lambda_j} \sigma_i
\sigma_j 
\end{align}
where we have defined $\bar m_{g} \equiv m_{g} {\cal{D}}^{1/2}/ h$. The dependence on the luminosity distance $D_L$ comes in implicitly as $z = z(A_0) = z(\mathcal{M}_z,D_L)$. Since the derivatives with respect to the model parameters are proportional to $\beta$ and evaluated at the injected parameters ($\beta=0$), Eq.~\eqref{eq:error_prop} reduces to
\begin{equation}
\label{eq:sigma-of-mg}
\sigma_{\bar{m}_{g}^{2}} = \frac{1 + z}{\pi^{2} \mathcal{M}_{z}} \sigma_{\beta}\,,
\end{equation} 
where we have used Eq.~\eqref{eq:beta_relation} to evaluate the derivative. From this expression, we can derive similar expressions for the variance of the Compton wavelength of the graviton, divided by the effective distance. Clearly, ${\cal{D}}$ (or equivalently $r_{\V}$) and $\lambda_{g}$ (or equivalently $m_{g}$) cannot be independently constrained due to the way they enter the $\beta$ parameter in Eq.~\eqref{eq:beta_relation}. 

\subsection{Waveform Model}

To produce the waveforms needed to compute the Fisher elements, we use the IMRPhenomD model by Khan, et. al.~\cite{Husa:2015iqa,Khan:2015jqa}. The model combines analytic, PN inspiral waveforms with phenomenological functions calibrated with numerical relativity simulations for the merger-ringdown phase. The fitting data spanned $\chi \in [-.95,.95]$ and $q = m_1/m_2 \leq 18$ and is designed to emulate spin-aligned or anti-aligned systems. The functional form of the waveform is  
\begin{equation}
\label{eq:generalwaveform}
\tilde{h}_{\GR}(f) = A_{\GR}(f)e^{i \varphi_{\GR}(f)} \,,
\end{equation}
where $A(f)$ and $\varphi(f)$ are piece-wise functions defined as
\begin{equation}\label{eq:imrpd_phase}
\varphi_{\GR}(f) = \begin{cases}
		\phi_{\ins} & f<0.018/M  \,, \\
		\phi_{\intt} & 0.018/M<f<0.5 f_{\RD} \,, \\
		\phi_{\mr} & 0.5 f_{\RD}<f \,, \\
	\end{cases}
\end{equation}
\begin{equation}\label{eq:imrpd_amp}
\begin{aligned}
A_{\GR}(f) &= \begin{cases}
	A_0 A_{\ins} & f< 0.014/M\,, \\
	A_0 A_{\intt} & 0.014/M < f < f_{\peak} \,, \\
	A_0 A_{\mr} & f_{\peak}< f \,,
 	\end{cases} \\
\end{aligned}
\end{equation}
with
\begin{equation}
A_0 = \sqrt{\frac{\pi}{30}} \frac{\mathcal{M}_z^2}{D_L} (\pi \mathcal{M}_z f)^{-7/6}\,.
\end{equation}
where $f_{\RD}$ and $f_{\peak}$ are the ringdown and peak frequencies, respectively. The specific functional forms for each piece of the waveform can be found in~\cite{Husa:2015iqa,Khan:2015jqa}. 

This model only reproduces GR waveforms, and to extend it to non-GR modifications, specifically to include $\beta$ [Eq.~\eqref{eq:beta_relation}] in the phase, we modify the model via 
\begin{equation}\label{eq:modified_imrphenomd}
\tilde{h}(f) = \tilde{h}_{\GR}(f) e^{i \beta u^{-1}} = A_{\GR}(f)e^{i (\varphi_{\GR}(f) + \beta u^{-1})}\,.
\end{equation}
Note that this modification enters in all phases of coalescence, because it is sourced by a correction to the dispersion relation, which is a \emph{propagation} effect and not a modification in the generation of the GWs. The modified waveform model depends on the eight parameters $\lambda^{a} = [\text{ln}A,\Phi_c,t_c, \text{ln} \mathcal{M}_z,\text{ln} \eta,\chi_s,\chi_a,\beta]$. 

As modifications to GR can come in through both generation and propagation effects generally, the veracity of our model might be questionable. However, it was shown in~\cite{Yunes:2016jcc} that the effects of modifications to the propagation of GWs overwhelm the effects due to the generation of GWs in general massive theories of gravity. This difference comes about because the propagation effect accumulates over the entire distance traveled, while the generation effects only develop while the system is producing GWs in band. Furthermore, confining our study to theories that exhibit Vainshtein screening should generally suppress any modification to the generation of gravitational waves. This is due to the fact that the scalar field is suppressed around the source, as was numerically verified in cubic Galileon by~\cite{Dar:2018dra,PhysRevD.87.044025}. These reasons lead us to believe that GW events in the modified theories we are examining are sufficiently similar to corresponding events in GR that calibration of our waveforms using GR numerical simulations is justified.

\section{Projected Constraints on Screened Massive Gravity}
\label{sec:proj-const}

In this section, we describe the projected constraints we find using a Fisher analysis for a variety of astrophysical parameters that are expected sources for the various ground-based and space-based detectors, as well as for the previously announced LIGO/VIRGO observations. The parameters of the chosen injections that we will study are shown in Table~\ref{table:parameter values}. We begin by presenting theory-agnostic, projected constraints, and then specialize to constraints on particular theories.   
\begin{table*}[htb]
\caption{Parameters of the models used in this study and the subsequent predicted bounds on $\beta$ for each model. All quantities are in the source frame. The bounds on $\beta$ are a 1$\sigma$ bound. The sources for aLIGO were picked to emulate previous detections by aLIGO/VIRGO~\cite{2041-8205-851-2-L35,Abbott:2018wiz,PhysRevLett.119.141101,Abbott:2017vtc,PhysRevLett.116.241103,PhysRevLett.116.241102} }
\label{table:parameter values}
\begin{center}
\begin{tabularx}{\textwidth}{ >{\centering\arraybackslash}X | >{\centering\arraybackslash}X >{\centering\arraybackslash}X >{\centering\arraybackslash}X >{\centering\arraybackslash}X >{\centering\arraybackslash}X >{\centering\arraybackslash}X >{\centering\arraybackslash}X}
\hline
Model & $M_1$ ($M_\odot$)& $M_2$ ($M_\odot$)& $\chi_1$ & $\chi_2$ & $D_L$ (Mpc)& SNR & $ \Delta \beta$  \\ \hline \hline
\multicolumn{8}{c}{aLIGO} \\ \hline
GW150914 & 36.0 & 29.0 & 0.32 & 0.44 & 410.0 &4.57 $\times 10^{ 1 }$ & 6.61 $\times 10^{ -2 }$ \\ [.5ex] \hline
GW151226 & 14.2 & 7.5 & 0.2 & 0.01 & 440.0 &1.64 $\times 10^{ 1 }$ & 7.75 $\times 10^{ -2 }$ \\ [.5ex] \hline
GW170104 & 31.2 & 19.4 & 0.45 & 0.47 & 880.0 &1.83 $\times 10^{ 1 }$ & 1.65 $\times 10^{ -1 }$ \\ [.5ex] \hline
GW170814 & 30.5 & 25.3 & 0.01 & 0.01 & 540.0 &2.88 $\times 10^{ 1 }$ & 7.20 $\times 10^{ -2 }$ \\ [.5ex] \hline
GW170817 & 1.48 & 1.26 & 0.02 & 0.02 & 44.7 &3.08 $\times 10^{ 1 }$ & 1.23 $\times 10^{ -1 }$ \\ [.5ex] \hline
GW170608 & 12.0 & 7.0 & 0.0 & 0.0 & 340.0 &1.89 $\times 10^{ 1 }$ & 7.77 $\times 10^{ -2 }$ \\ [.5ex] \hline
\hline

\multicolumn{8}{c}{Einstein Telescope} \\ \hline
1 & 1.4 & 1.0 & 0.03 & 0.01 & 50.0 &3.55 $\times 10^{ 2 }$ & 2.93 $\times 10^{ -3 }$ \\ [.5ex] \hline
2 & 500.0 & 100.0 & 0.1 & 0.3 & 1500.0 &5.47 $\times 10^{ 2 }$ & 3.84 $\times 10^{ -3 }$ \\ [.5ex] \hline
3 & 70.0 & 50.0 & 0.7 & 0.9 & 400.0 &1.22 $\times 10^{ 3 }$ & 7.19 $\times 10^{ -4 }$ \\ [.5ex] \hline
4 & 50.0 & 3.0 & 0.7 & 0.7 & 400.0 &2.63 $\times 10^{ 2 }$ & 6.80 $\times 10^{ -4 }$ \\ [.5ex] \hline
5 & 80.0 & 40.0 & 0.9 & 0.2 & 600.0 &7.77 $\times 10^{ 2 }$ & 1.35 $\times 10^{ -3 }$ \\ [.5ex] \hline
6 & 100.0 & 40.0 & 0.4 & 0.3 & 2000.0 &2.63 $\times 10^{ 2 }$ & 1.02 $\times 10^{ -2 }$ \\ [.5ex] \hline
\hline

\multicolumn{8}{c}{LISA} \\ \hline
7 & 6 $\times 10^{ 6 }$ & 5 $\times 10^{ 6 }$ & 0.32 & 0.44 & 1.00 $\times 10^{ 4 }$ &3.31 $\times 10^{ 3 }$ & 1.23 $\times 10^{ -3 }$ \\ [.5ex] \hline
8 & 6 $\times 10^{ 6 }$ & 5 $\times 10^{ 6 }$ & 0.7 & 0.8 & 3.00 $\times 10^{ 4 }$ &9.21 $\times 10^{ 2 }$ & 7.38 $\times 10^{ -3 }$ \\ [.5ex] \hline
9 & 5 $\times 10^{ 6 }$ & 4 $\times 10^{ 5 }$ & 0.45 & 0.47 & 8.80 $\times 10^{ 3 }$ &1.26 $\times 10^{ 3 }$ & 1.16 $\times 10^{ -3 }$ \\ [.5ex] \hline
10 & 5 $\times 10^{ 5 }$ & 4 $\times 10^{ 4 }$ & 0.01 & 0.01 & 5.40 $\times 10^{ 3 }$ &1.57 $\times 10^{ 3 }$ & 3.72 $\times 10^{ -4 }$ \\ [.5ex] \hline
11 & 5 $\times 10^{ 4 }$ & 4 $\times 10^{ 3 }$ & 0.7 & 0.9 & 1.60 $\times 10^{ 4 }$ &1.99 $\times 10^{ 2 }$ & 9.80 $\times 10^{ -4 }$ \\ [.5ex] \hline
12 & 5 $\times 10^{ 6 }$ & 4 $\times 10^{ 6 }$ & 0.7 & 0.9 & 4.80 $\times 10^{ 4 }$ &5.32 $\times 10^{ 2 }$ & 1.39 $\times 10^{ -2 }$ \\ [.5ex] \hline
\end{tabularx}
\end{center}
\end{table*}

\subsection{Theory Agnostic}

The constraints on $\beta$ for each of the injections we considered are listed in the last column of Table~\ref{table:parameter values}. We can then map these constraints to bounds on $r_{\V}$ and $\lambda_{g}$ (or $m_{g}$) through Eq.~\eqref{eq:sigma-of-mg}, leaving $r_{\V}$ as an independent parameter. This allows for the results to be interpreted in the context of any theory that exhibits Vainshtein screening with a graviton mass. The Vainshtein radius, however, generally depends on the Schwarzschild radius of the Milky Way and that of the host galaxy, in principle creating an asymmetric screening scenario if the host galaxy mass differs substantially from the mass of the Milky Way. To simplify the analysis below, we assume the host galaxy mass is comparable to that of the Milky Way, therefore creating a symmetric system.  

\begin{figure}[htb!]
\hspace*{-0.6cm}\includegraphics[width=9.75cm,clip=true]{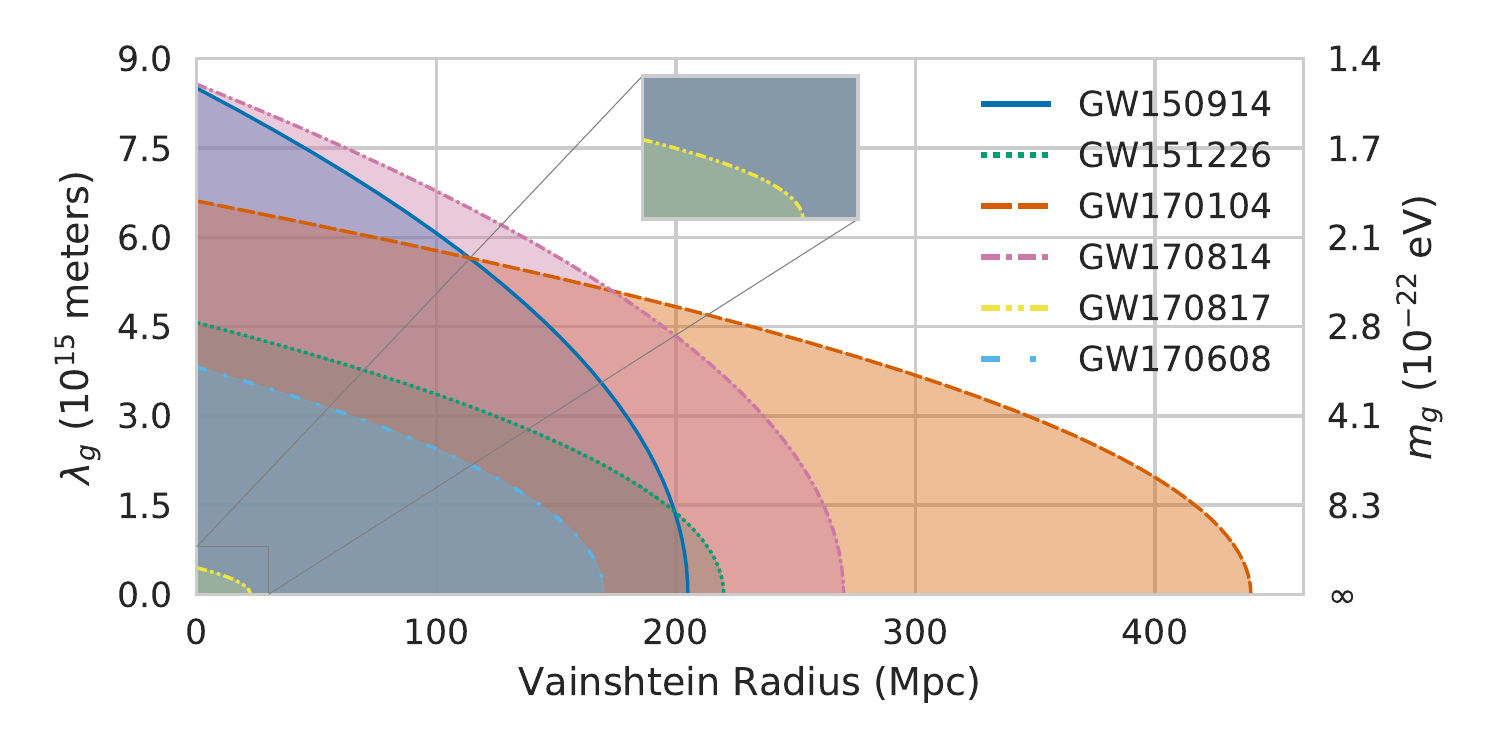}
\hspace*{-0.6cm}\includegraphics[width=9.75cm,clip=true]{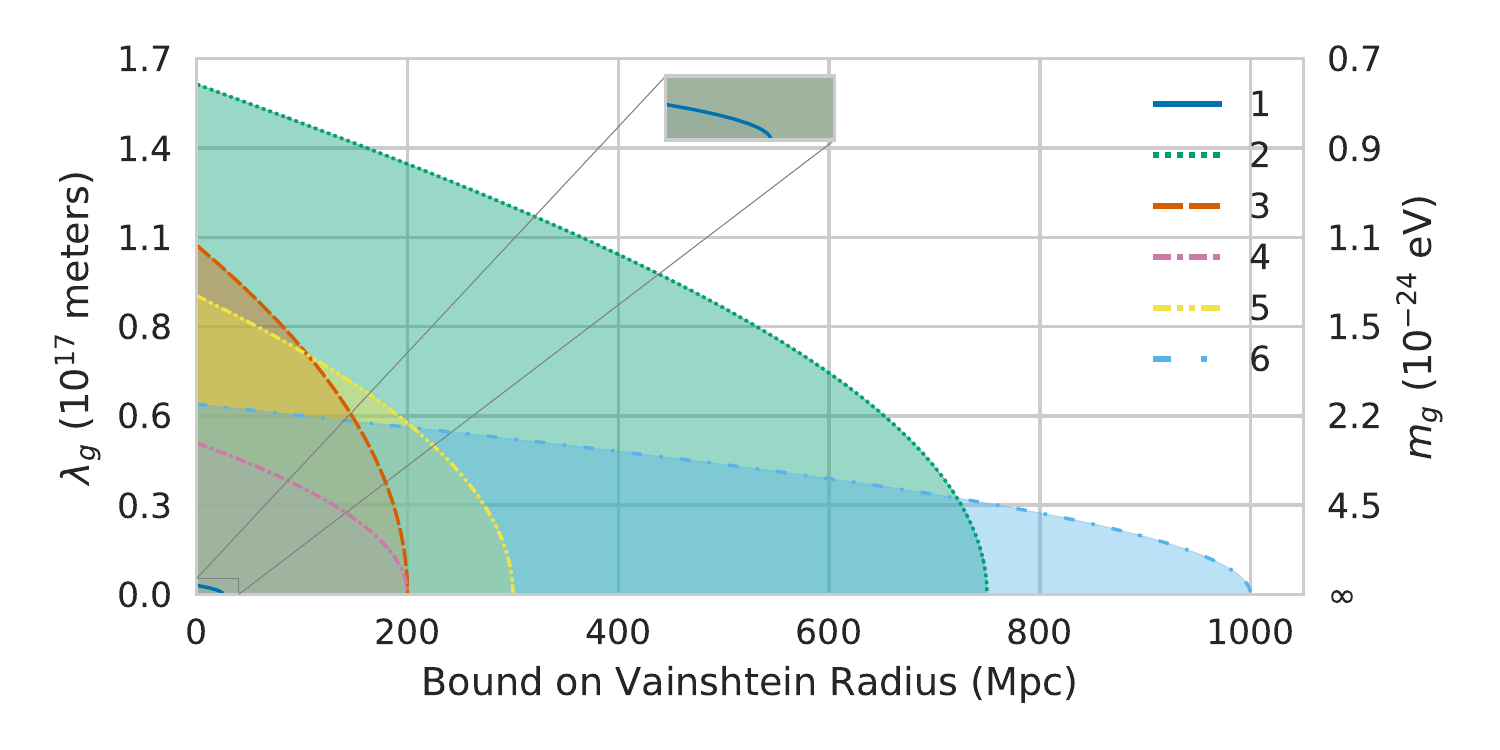}
\hspace*{-0.6cm}\includegraphics[width=9.75cm,clip=true]{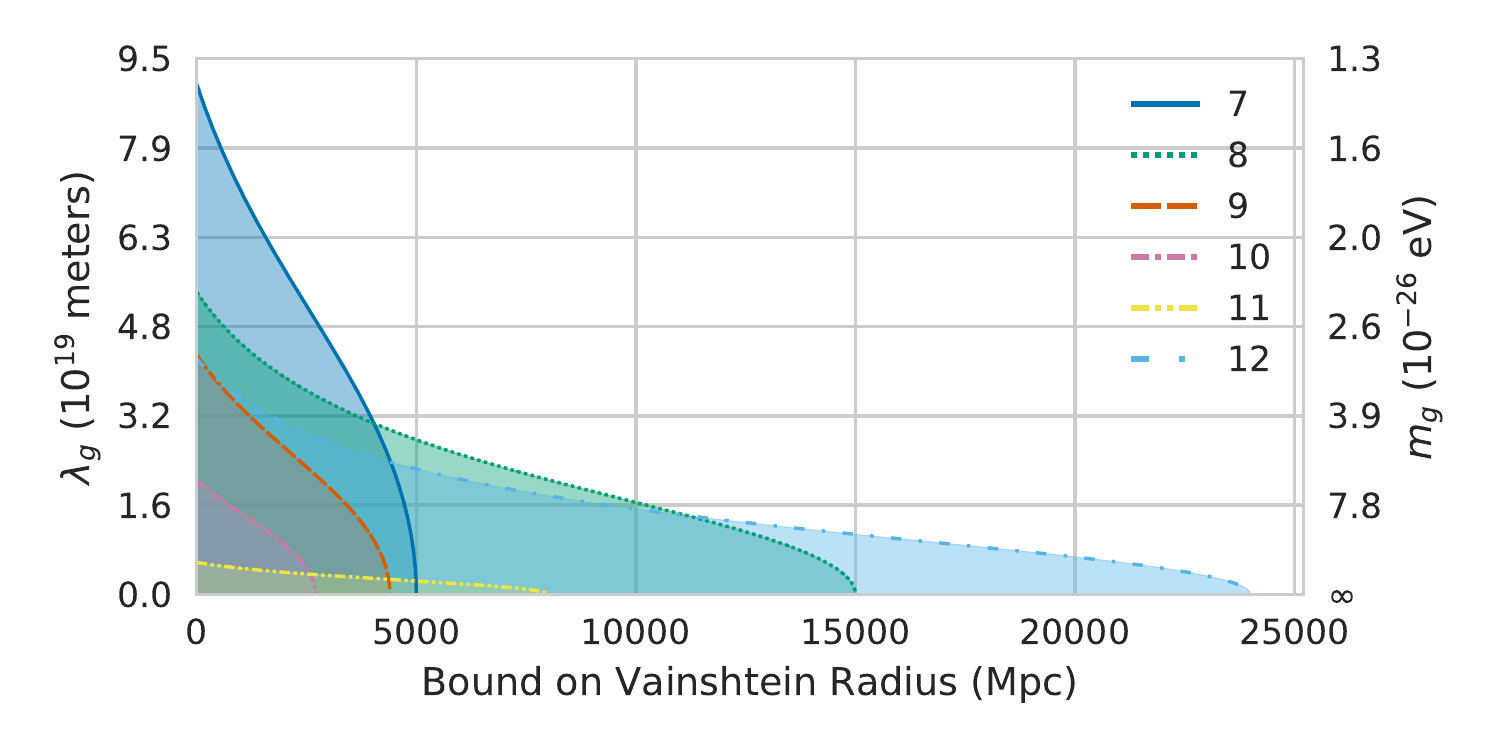}
\caption{(Color Online) Projected constraint on the mass of the graviton as a function of the Vainshtein radius for a variety of systems detected with aLIGO (top), ET (middle) and LISA (bottom). The shaded regions would be disallowed given injections consistent with GR. }
\label{fig:observations}
\end{figure}

The projected constraints in the $m_{g}$--$r_{\V}$ (or $\lambda_{g}$--$r_{\V}$) plane are shown in Fig.~\ref{fig:observations} for aLIGO, LISA, and ET. 
For aLIGO sources, the projected constraints follow a uniform pattern. For each source, the effect of screening increasingly grows more pronounced until reaching the midpoint between the source and the Milky Way, corresponding to a fully screened graviton mass. Systems with a higher mass and at closer distances allow for stronger constraints on $\beta$ (and therefore on $m_g$ and $r_\V$), but systems at much higher redshifts allow for a deeper probing of the screening mechanism at much larger distances from screening sources. The low SNR, low chirp mass of NSNS binaries conspire to inhibit our ability to constrain $m_g$, and as such, are poor sources to constrain these effects.

The prospects of probing screened massive gravity improve dramatically when incorporating potential LISA and ET detections, as these detectors will be able to observe sources much farther out and with a higher SNR. Due to the nonlinear relation between the cosmological distance in Eq.~\eqref{eq:defD} and the luminosity distance, not all distances are equal for this type of test. Sources farther away are not just rescaled versions of closer sources, but actually probe more complex domains of the $m_g$--$r_\V$ relation. The abrupt change in our ability to constrain the graviton mass gives way to a much softer effect at farther screening radii where large changes in screening radius impact our constraints much less. The development of third generation detectors, like ET and LISA, will also allow for a large population of observations, which could also be leveraged to improve constraints via stacking. As the redshift grows, these constraints become increasingly sensitive to the cosmological model employed because of the dependence of $\cal{D}$ on the Hubble function $H(z)$.

\subsection{dRGT and bigravity}\label{section:drgt_bigravity}

Let us now consider projected constraints on specific modified gravity models that predict a massive graviton and screening. In the context of specific theories like dRGT and bigravity, the screening radius becomes a function of the graviton mass. Having analytic expressions for the screening radius is useful for exploring the possible bounds on these parameters, but it should be noted that the theories themselves do not necessarily realize these types of modifications~\cite{deRham:2016nuf,deRham:2014zqa}. Replacing the parameter $r_\V$ with the relation shown in Eq.~\eqref{eq:drgtradius} analytically breaks the degeneracy between $r_{\V}$ and $m_{g}$. To calculate the graviton mass constraint, as the relationship between $\cal{D}$ and $r_\V$ is highly nonlinear, we numerically solve Eq.~\eqref{eq:beta_relation}. The results are shown in Table~\ref{table:drgtbounds}, along with the corresponding screening radius for a Milky Way sized galaxy and the percent change in the graviton mass bound if screening were not incorporated.

In the context of dRGT- and bigravity-type screening radii, screening has a minimal effect given the projected sensitivity of second- and third-generation detectors. This is because saturating the bounds on $m_{g}$ gives screening radii of $\mathcal{O}(1 \text{pc})$, since $r_\V$ is proportional to $1/m_g^2$. Once detectors become sensitive enough to begin probing the graviton mass at the $\mathcal{O}(H_0)$ level, the screening radius for a Milky Way sized galaxy will grow to $\mathcal{O}(\text{Mpc})$. With a screening radius of this magnitude, the effects might warrant further investigation. 

Examining Eq.~\eqref{eq:beta_relation}, we see that the ideal candidate source to place constraints will have a large chirp mass and a low redshift, as screening effects encompassed by $\cal{D}$ are enhanced and degraded by high chirp mass and high redshift, respectively. As space-based detectors will target this higher range of chirp masses, sources being targeted by future LISA-type missions, like the Big Bang Observer~\cite{PhysRevD.72.083005}, would be the ideal subjects of future studies on this mechanism. Even though the average redshift of these millihertz systems will be higher, the average chirp mass will be many orders of magnitude larger than targeted sources for current and future ground-based detectors, overwhelming any loss due to higher distance. 

\begin{table}[htb]
\begin{tabular}{  ||>{\centering\arraybackslash}m{2cm} | >{\centering\arraybackslash}m{2cm} >{\centering\arraybackslash}m{2cm}  >{\centering\arraybackslash}m{2cm} || }
\hline
Model  & $\Delta m_g$ (Screened) & Percent Change (\%)  & $r_\V$ \\ \hline \hline
\multicolumn{4}{|c|}{aLIGO} \\ \hline
GW150914 & 1.46 $\times 10^{ -22 }$ & 0.03 & 3.76 $\times 10^{ -8 }$ \\ [.5ex] \hline
GW151226 & 2.72 $\times 10^{ -22 }$ & 0.03 & 2.49 $\times 10^{ -8 }$ \\ [.5ex] \hline
GW170104 & 1.88 $\times 10^{ -22 }$ & 0.06 & 3.18 $\times 10^{ -8 }$ \\ [.5ex] \hline
GW170814 & 1.45 $\times 10^{ -22 }$ & 0.04 & 3.78 $\times 10^{ -8 }$ \\ [.5ex] \hline
GW170817 & 2.82 $\times 10^{ -21 }$ & 0.0 & 5.22 $\times 10^{ -9 }$ \\ [.5ex] \hline
GW170608 & 3.25 $\times 10^{ -22 }$ & 0.02 & 2.21 $\times 10^{ -8 }$ \\ [.5ex] \hline
\hline

\multicolumn{4}{|c|}{ET} \\ \hline
1 & 4.44 $\times 10^{ -22 }$ & 0.0 & 1.79 $\times 10^{ -8 }$  \\ [.5ex] \hline
2 & 7.79 $\times 10^{ -24 }$ & 0.09 & 2.65 $\times 10^{ -7 }$  \\ [.5ex] \hline
3 & 1.14 $\times 10^{ -23 }$ & 0.03 & 2.06 $\times 10^{ -7 }$  \\ [.5ex] \hline
4 & 2.63 $\times 10^{ -23 }$ & 0.03 & 1.18 $\times 10^{ -7 }$  \\ [.5ex] \hline
5 & 1.33 $\times 10^{ -23 }$ & 0.04 & 1.86 $\times 10^{ -7 }$  \\ [.5ex] \hline
6 & 2.09 $\times 10^{ -23 }$ & 0.1 & 1.37 $\times 10^{ -7 }$  \\ [.5ex] \hline
\hline

\multicolumn{4}{|c|}{LISA} \\ \hline
7 & 1.37 $\times 10^{ -26 }$ & 0.24 & 1.82 $\times 10^{ -5 }$  \\ [.5ex] \hline
8 & 2.31 $\times 10^{ -26 }$ & 0.29 & 1.29 $\times 10^{ -5 }$  \\ [.5ex] \hline
9 & 2.92 $\times 10^{ -26 }$ & 0.23 & 1.10 $\times 10^{ -5 }$  \\ [.5ex] \hline
10 & 6.17 $\times 10^{ -26 }$ & 0.19 & 6.68 $\times 10^{ -6 }$  \\ [.5ex] \hline
11 & 2.19 $\times 10^{ -25 }$ & 0.27 & 2.87 $\times 10^{ -6 }$  \\ [.5ex] \hline
12 & 2.97 $\times 10^{ -26 }$ & 0.3 & 1.09 $\times 10^{ -5 }$  \\ [.5ex] \hline
\end{tabular}
\caption{\label{table:drgtbounds} Projected $1\sigma$ bounds for $m_g$ for dRGT- and bigravity-type screening radii. The second column shows the percent change in the constraint on $m_g$ between the screened and unscreened constraints, while the screening radius $r_\V$ is measured in Mpc, and the graviton mass in eV.}
\end{table}

\section{Conclusions and Future Work}
\label{sec:conclusions}

We studied the effect of screening on projected, future constraints on the mass of the graviton from a model-agnostic viewpoint and then within a model-specific approach. When studying model-independent constraints, we found that the effect of screening is to generically deteriorate bounds on the graviton mass. The reason for this is that screening eliminates the correction to the dispersion relation during part of the graviton travel, thus decreasing the overall correction to the GW phase. When studying model-dependent constraints, we found that the effect of screening on graviton mass bounds is actually quite small for the models considered. The reason for this is that the relation between the screening radius and the mass of the graviton is such that, for the range of graviton masses that second- and third-generation detectors are sensitive to, the screening radius is very small.  

Future work could focus on a variety of topics. As we enter the era of third-generation detectors, or even second-generation detectors at design sensitivity, the number of GW observations will greatly increase. Constraints on the mass of the graviton, therefore, can be enhanced by combining multiple observations. Doing so can have interesting effects on model-agnostic constraints on the screening radius. In the model-dependent case, stacked constraints could become enough that the effect of screening is not negligible and ultimately ought to be taken into account. 

Another possible avenue for future work could involve the more careful study of screening within known models. As we described in this paper, Vainshtein screening has really only been studied in detail within some greatly simplified physical models, such as in stationarity and spherical symmetry. The compact binaries that generate the GWs that we detect, however, are neither stationary, nor spherically symmetric. The degree of effectiveness of screening in such cases should be investigated further. 

A final possibility for future work could consist on the study of other cosmological modified theories to determine whether they also predict screening. Theories such as bigravity, or Horndeski and beyond Horndeski theories could be studied further to determine formally whether the dispersion of GWs is partially screened in regions of high density or high curvature. It is possible that screening is not present for GWs in some of these theories, as is the case in quartic and quintic Galileon theories, but this ought to be demonstrated with careful calculation.    

\acknowledgments
The authors would like to thank Claudia de Rham, Rachel Rosen, Hector Okada da Silva, and Mark Trodden for useful conversations and input. We would also like to thank Kent Yagi for comparisons with his Fisher code. The authors acknowledge support from NSF grant PHY-1759615 and NASA grants NNX16AB98G and 80NSSC17M0041.

\bibliographystyle{apsrev4-1}
\bibliography{paper}

\end{document}